\documentclass[pdflatex,sn-mathphys-num]{sn-jnl}

\usepackage{graphicx}%
\usepackage{multirow}%
\usepackage{amsmath,amssymb,amsfonts}%
\usepackage{amsthm}%
\usepackage{mathrsfs}%
\usepackage[title]{appendix}%
\usepackage{xcolor}%
\usepackage{textcomp}%
\usepackage{manyfoot}%
\usepackage{booktabs}%
\usepackage{algorithm}%
\usepackage{algorithmicx}%
\usepackage{algpseudocode}%
\usepackage{listings}%
\usepackage[utf8]{inputenc}
\usepackage{graphicx} 
\usepackage{soul}
\usepackage{hyperref}
\hypersetup{
  colorlinks,%
  citecolor=black,%
  filecolor=black,%
  linkcolor=black,%
  urlcolor=blue
  }

\begin{document}

\title{Revisiting the soft-hard separation in the transverse momentum spectra of \textit{pp} collisions.} 

\author*[1,2]{\fnm{G\'abor} \sur{B\'ir\'o}}\email{biro.gabor@wigner.hun-ren.hu}

\author[3]{\fnm{Guy} \sur{Pai\'c}}\email{Guy.Paic@cern.ch}

\author[4]{\fnm{Leonid} \sur{Serkin}}\email{lserkin@ciencias.unam.mx}

\author[1]{\fnm{Gergely G\'abor} \sur{Barnaf\"oldi}}\email{barnafoldi.gergely@wigner.hun-ren.hu}

\affil[1]{\orgname{HUN-REN Wigner Research Center for Physics},
\orgaddress{\street{29--33 Konkoly-Thege Mikl\'os Str.}, \city{Budapest},
\postcode{H-1121}, \country{Hungary}}}

\affil[2]{\orgname{ELTE E\"otv\"os Lor\'and University, Institute of Physics},
\orgaddress{\street{1/A P\'azm\'any P\'eter S\'et\'any}, \city{Budapest},
\postcode{H-1117}, \country{Hungary}}}

\affil[3]{\orgname{Instituto de Ciencias Nucleares, Universidad Nacional Aut\'onoma de M\'exico},
\orgaddress{\street{Apartado Postal 70-543}, \city{Ciudad de M\'exico},
\postcode{04510}, \country{M\'exico}}}

\affil[4]{\orgname{Facultad de Ciencias, Universidad Nacional Aut\'onoma de M\'exico},
\orgaddress{\street{Circuito Exterior s/n}, \city{Ciudad de M\'exico},
\postcode{04510}, \country{M\'exico}}}

\date{\today}

\abstract{We study the separation of soft and hard components in the transverse momentum spectra of charged particles as measured by ALICE in proton-proton collisions at {\unboldmath$\sqrt{s}$} = 2.76, 5.02 and 13~TeV at the LHC. The soft component is described by a Boltzmann fit, while the residual spectra are identified as a hard QCD-like fragmentation contribution. After separation, the subtracted spectra show no significant evolution in shape or peak position with multiplicity, consistent with a two-component interpretation. Mean transverse momenta for both contributions remain nearly constant across multiplicity classes, while {\sc Pythia}~8 Monte Carlo simulations confirm these trends. The robustness of the decomposition is demonstrated by a comparison between simulations with and without color reconnection, yielding consistent results. The terms `soft' and `hard' are used as operational labels within this framework and should not be interpreted as uniquely identified dynamical components. Our results are consistent with the two-component description as a viable and physically motivated alternative to hydrodynamical interpretations.}

\maketitle

\section{Introduction}
Hadron production in high-energy nuclear collisions has been extensively studied since the 1950s, with dozens of experimental collaborations formed at state-of-the-art particle accelerators~\cite{Evans:2008zzb}. To understand the nature of the strong interaction, Quantum Chromodynamics (QCD) was developed and used effectively at the highest transverse momentum ($p_T$) regimes where fragmentation plays a role in hadron formation (see e.g. \cite{ALICE:2022wpn} and references therein). In contrast, the low-$p_T$ behaviour of the spectra, belonging to the soft sector, still lacks a theory based on first principles, although many phenomenological approaches remain quite successful and show strong predictive power.

The interpretation of the hadron spectra measured at different energies has been a source of controversy, stemming from the Janus-faced nature of the strong interaction. This understanding is further complicated by the observation that strong collective effects appear across the full range of colliding systems, from small to large sizes~\cite{STRICKLAND201992}. In general terms, one can follow essentially two approaches. The first is the two-component picture, based on numerous attempts to separate the soft and hard parts of the measured spectra. Under the assumption that there are indeed well-defined processes: on one hand, the so-called ``soft'' processes, which are theoretically not amenable to precise simulation (see for instance~\cite{UA2:1987hpe, Wang:1988bw, Ceradini:1985vm}); and on the other, the ``hard'' processes of QCD nature.

It is important to note that, despite the dominant view embracing hydrodynamical interpretations, several works have consistently challenged this perspective and emphasized the need to account for the two-component nature of the spectra~\cite{Trainor:2018dfp, Trainor:2018hkw, Trainor:2022pob}. The spectra, in this interpretation, represent a sum of two distinct contributions: one arising from soft, nonperturbative QCD interactions, and the other from hard, perturbative QCD processes. The second component is represented in the hadron spectra as debris resulting from the fragmentation of partons in the hard interactions. 

The complexity arises from the fact that the fragmentation functions of jets produce a significant number of particles at relatively low transverse momenta. The superimposition of the two mechanisms in the low-$p_T$ region causes significant difficulties in the analysis of the data~\cite{CDF:2001hmt}. On the other extreme, since the first heavy ion experiments at SPS, it has been necessary to interpret the spectra of charged particles as originating from a thermal distribution subjected to isotropic expansion, driven by collective flow~(\cite{NA44:1998hna, Roland:1999zby, Andersen:1998vu, ATLAS:2025ztg} and references therein), usually accompanied by a theoretical treatment called ``the Blast Wave'' (BW). At higher collision energies, first at RHIC~\cite{STAR:2008med} and later at the LHC, the initial results on proton-proton ($pp$) collisions led to the application of the BW model to the data, giving rise to a large list of results claiming the existence of flow even in such light systems~\cite{Kisiel:2010xy, Florkowski:2016xbk, Biro2025}.

A parametrization over the full $p_T$ range can be achieved rather well with the Tsallis\,--\,Pareto function~\cite{Tsallis:1987eu}, and it has therefore been widely used in the literature for $pp$ and heavy-ion collisions~\cite{Gu:2022xjn, Biro:2020kve}. This success, however, comes with caveats. The Tsallis-distribution form can be derived from the non-additive Tsallis entropy form, which is a generalization of the Boltzmann\,--\,Gibbs statistics with strongly coupled cross term. This term can be phenomenologically associated with QCD-like behaviour presenting a common physical description of the soft and hard contributions. It provides an effective fit to the entire spectrum with a small number of parameters. In practice, this means that while the Tsallis\,--\,Pareto distribution can reproduce the shape of the spectra over a broad $p_T$ interval, it may obscure the underlying mechanisms, mixing fragmentation and soft production into a single effective temperature-like parameter for the distribution body and a power-like, non-extensivity parameter for the tail~\cite{Yang:2021bab}. In particular, since the Tsallis--Pareto form applies over the full $p_T$ range, it cannot serve as a clean soft-component model for the purposes of a two-component decomposition.


It is also worth noting that the blast-wave model, while successful in describing spectra in heavy-ion collisions, was designed under the assumption of collective radial flow. Applying it to isolate the ``soft'' component in $pp$ collisions would implicitly assume the very hypothesis to be tested. A more neutral and minimal soft ansatz is therefore preferred.

It is clear that the two interpretations, namely the existence of two distinct components in the spectra and the blast-wave approach, lead to fundamentally different conclusions. In this paper we aim to demonstrate that the experimental data are consistent with the two-component framework, which offers a qualitatively different perspective on the behaviour of hadron spectra in proton-proton collisions.

\section{Disentangling soft and hard contributions in the $\textrm{\boldmath{$p_T$}}$ spectra}

We analyse ALICE collision data taken at $\sqrt{s} = 13$ TeV in $pp$ collisions at the LHC~\cite{ALICE:2020nkc}, focusing on the $p_T$-spectra in different charged particle multiplicity classes as measured by the experiment in the central rapidity region. Our aim is to investigate whether the spectra can be described by separating the soft and hard contributions using the minimal set of assumptions.

We emphasize that throughout this analysis the terms ``soft'' and ``hard'' refer to \emph{operational definitions} within our two-component framework: the ``soft'' component is the yield described by the Boltzmann parametrization in the low-$p_T$ region ($p_T < p_0$), and the ``hard'' (fragmentation) component is the residual after subtraction. These are not claims about unique dynamical mechanisms, but convenient labels for the two contributions. In the transition region near $p_0$, contributions from both mechanisms are present and cannot be cleanly separated.

We approach the soft part of the ALICE data $p_T$ spectra using a very simple analytical form. A Boltzmann function was chosen to describe the shape of the spectra in the lowest momentum region, where soft production dominates. This part of the spectra is known to be a complicated mix of non-perturbative effects such as multiple parton interactions (MPI), intrinsic transverse momentum, and initial or final state radiation. The soft component, parametrized by the Boltzmann exponential, is associated with these low-momentum-transfer non-perturbative processes -- including MPI at low virtuality, soft string fragmentation, and underlying-event activity -- that produce the bulk of particles at $p_T \lesssim 1$~GeV/$c$. The hard component, obtained as the residual after subtraction, is associated with particles originating from the fragmentation of partons produced in hard ($Q^2 \gg \Lambda_{\rm QCD}^2$) perturbative scatterings, which at high $p_T$ follow a power-law distribution characteristic of pQCD fragmentation functions. Our goal here was not to assign a physical temperature but to test how well the Boltzmann form can represent the data in the low-$p_T$ region. Accordingly, we have fitted the lowest part of the spectra with the following function:
\begin{equation}
    f(p_T)=A\exp\left[-\beta p_T\right] \  ,
    \label{eq:boltzmann}
\end{equation}
where $\beta=1/T$ is commonly interpreted as the slope or the inverse effective temperature, $T$, and $A$ is a normalization constant.

We note that the decomposition is inherently dependent on the choice of the soft parametrization. The Boltzmann function is the simplest two-parameter ansatz applicable in the restricted low-$p_T$ regime. The conclusions regarding the multiplicity independence of the hard component's shape are not expected to be strongly sensitive to this choice within the restricted fit range below $p_0$.

\begin{figure}[t]
\centering
\includegraphics[width=0.48\linewidth]{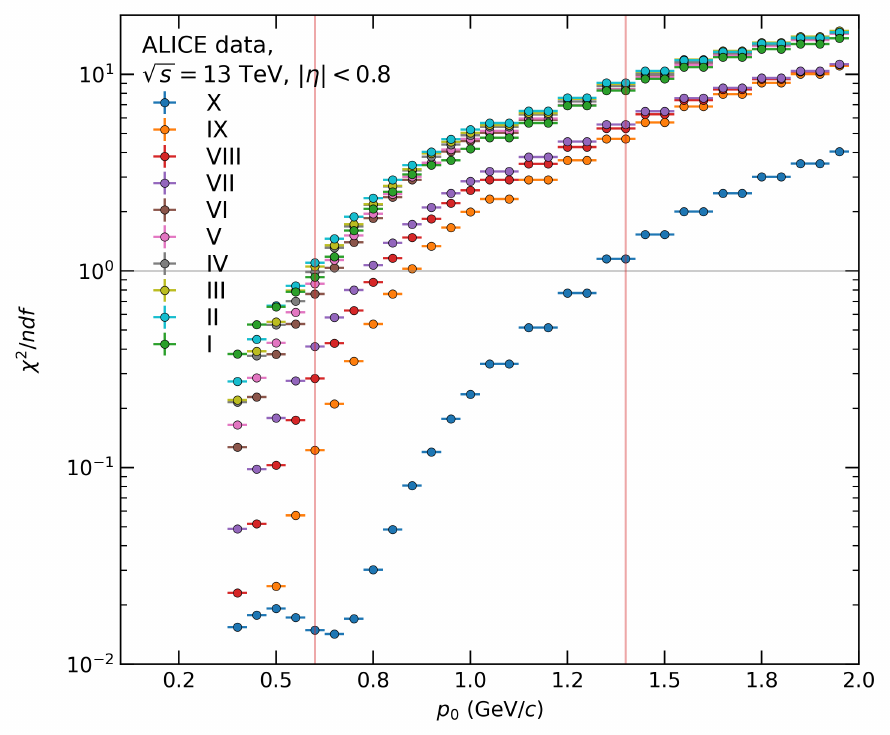}
\caption{
The $\chi^2/\mathrm{ndf}$ values of the Boltzmann fits with respect to the cut parameter $p_0$ for ALICE pp collisions at $\sqrt{s} = 13$ TeV, shown for various multiplicity classes~\cite{ALICE:2020nkc}. Vertical lines show the $p_0$ values for the highest and lowest multiplicities.
}
\label{fig:chi2ndf}
\end{figure}

We classified the events by multiplicity, and at each multiplicity we separated the soft and hard parts of the $p_T$ spectra by fitting the Boltzmann function up to a certain transverse momentum, denoted by $p_0$. This parameter defines the upper limit of the fit and marks the region where the contribution from hard QCD processes is still negligible. The optimal $p_0$ was chosen as the value where the $\chi^2/\text{ndf}$ is closest to one (see Fig.~\ref{fig:chi2ndf}). This procedure was repeated for each spectrum measured by ALICE at 13~TeV~\cite{ALICE:2020nkc}. One observes that the $\chi^2/\text{ndf}$ values show a strong dependence on multiplicity. This is expected, since at low multiplicity the contribution from fragmentation is minimal, and the soft component dominates.

The parameter $p_0$ should be understood as a data-driven, analysis-dependent upper limit of the soft fit region, determined by an objective $\chi^2/\text{ndf}$ criterion. It is not directly derived from a first-principles QCD scale, but the values obtained ($p_0 \sim 0.9$--$1.1$~GeV/$c$) are consistent with the phenomenological expectation of a soft--hard transition in the vicinity of scales at which non-perturbative QCD effects become dominant. The multiplicity dependence of $p_0$ reflects the fact that at higher multiplicities the contribution from hard scatterings starts to distort the spectrum at lower $p_T$, consistent with an increased rate of hard processes.

\begin{figure}[b]
\centering
\includegraphics[width=0.48\linewidth]{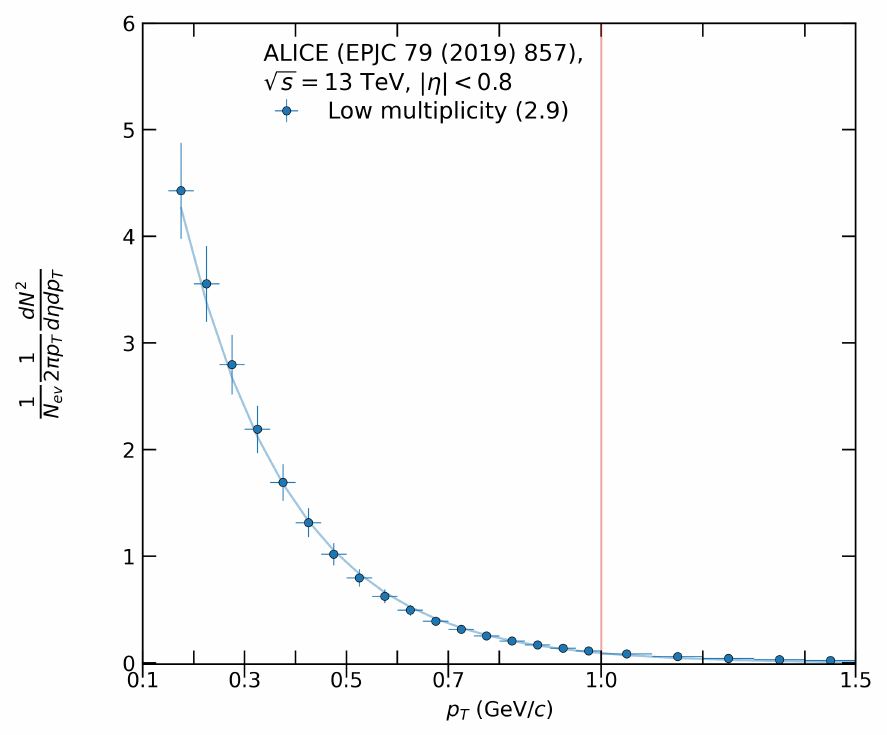}
\includegraphics[width=0.48\linewidth]{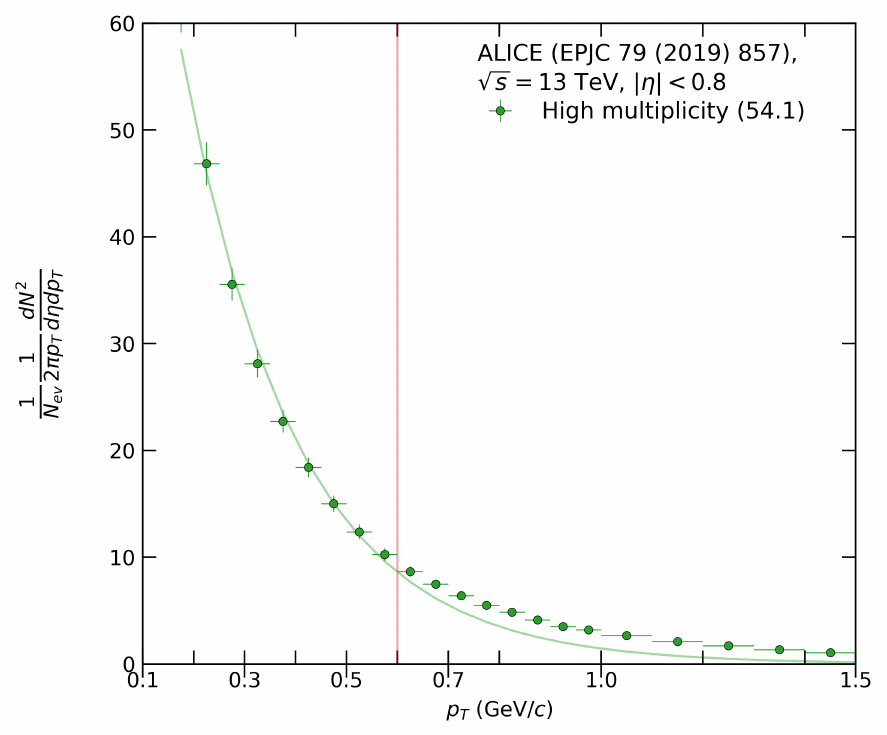}
\caption{
Boltzmann-distribution fits shown on linear scale at $\sqrt{s} = 13$~TeV in pp collisions, for ALICE data in the (top) lowest and (bottom) highest multiplicity classes~\cite{ALICE:2020nkc}. Vertical lines indicate the obtained $p_0$ values for which $\chi^2/\mathrm{ndf} < 1.0$.
}
\label{fig:fitlinear}
\end{figure}

To assess the performance of the fits, the resulting spectra are shown in linear scale in Fig.~\ref{fig:fitlinear} for the two extreme multiplicity classes. As seen in Fig.~\ref{fig:chi2ndf}, the hadron spectra in the lowest multiplicity bin ($N_{\text{ch}} \sim 2.9$) are well described by the Boltzmann function up to $p_T \gtrsim 1.1$~GeV/$c$, or even higher. At high multiplicity ($N_{\text{ch}} \sim 54.1$), however, one observes a gradual worsening of the fit already at lower $p_0$ values, as the contribution from fragmentation becomes increasingly significant.

\begin{figure}[h]
\centering
\includegraphics[width=0.48\linewidth]{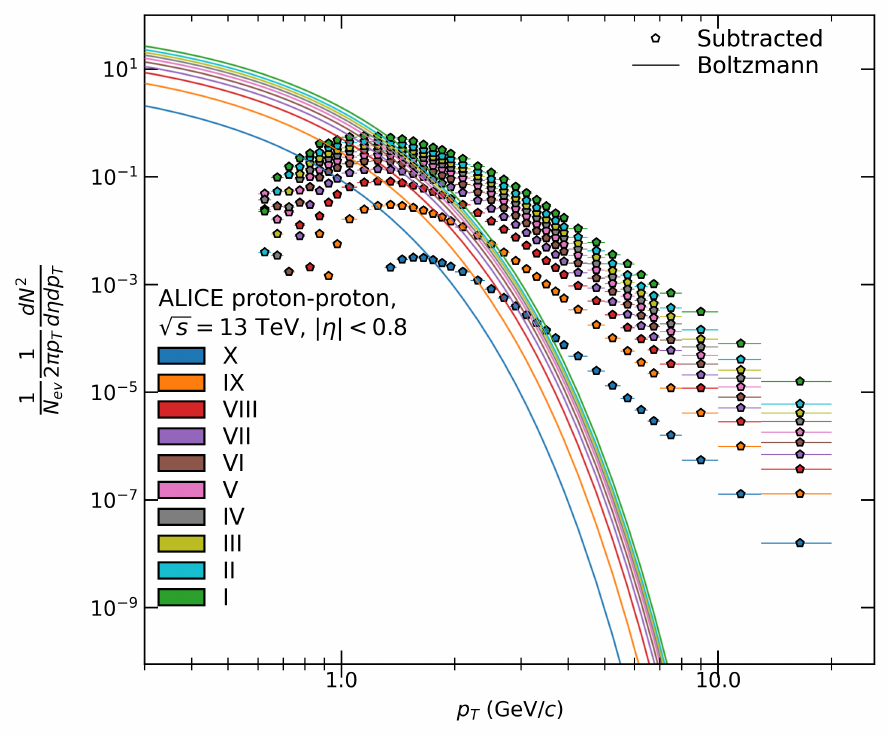}
\caption{
Subtracted hadron spectra obtained by removing the fitted Boltzmann component from the ALICE data~\cite{ALICE:2020nkc}. The remaining spectra represent the fragmentation contribution in different multiplicity classes. Solid lines correspond to the original Boltzmann fits.}
\label{fig:subtracted}
\end{figure}

The next step is to subtract the Boltzmann fits, representing the soft contributions, from the total spectra. The resulting distributions then represent the fragmentation components. As seen in Fig.~\ref{fig:subtracted}, beyond the $p_0$ value, the spectra still carry some contribution from the soft QCD component. Moreover, the deviation from the expected power-law behaviour of hard QCD is clearly visible in the mismatch between the exponential and power-law regions. The result is interesting: one observes that neither the shape of the Boltzmann component nor that of the high-$p_T$ part, attributed to fragmentation, shows any significant evolution in the position of their peaks with multiplicity. This observation is consistent with the absence of an additional intermediate source such as radial flow or recombination, within the context of this decomposition, though it does not uniquely rule out such contributions. The two-component description, consisting of a soft Boltzmann and a hard QCD-like part, seems sufficient to describe the data across multiplicity classes.

\begin{figure}[h]
\centering
\includegraphics[width=0.48\linewidth]{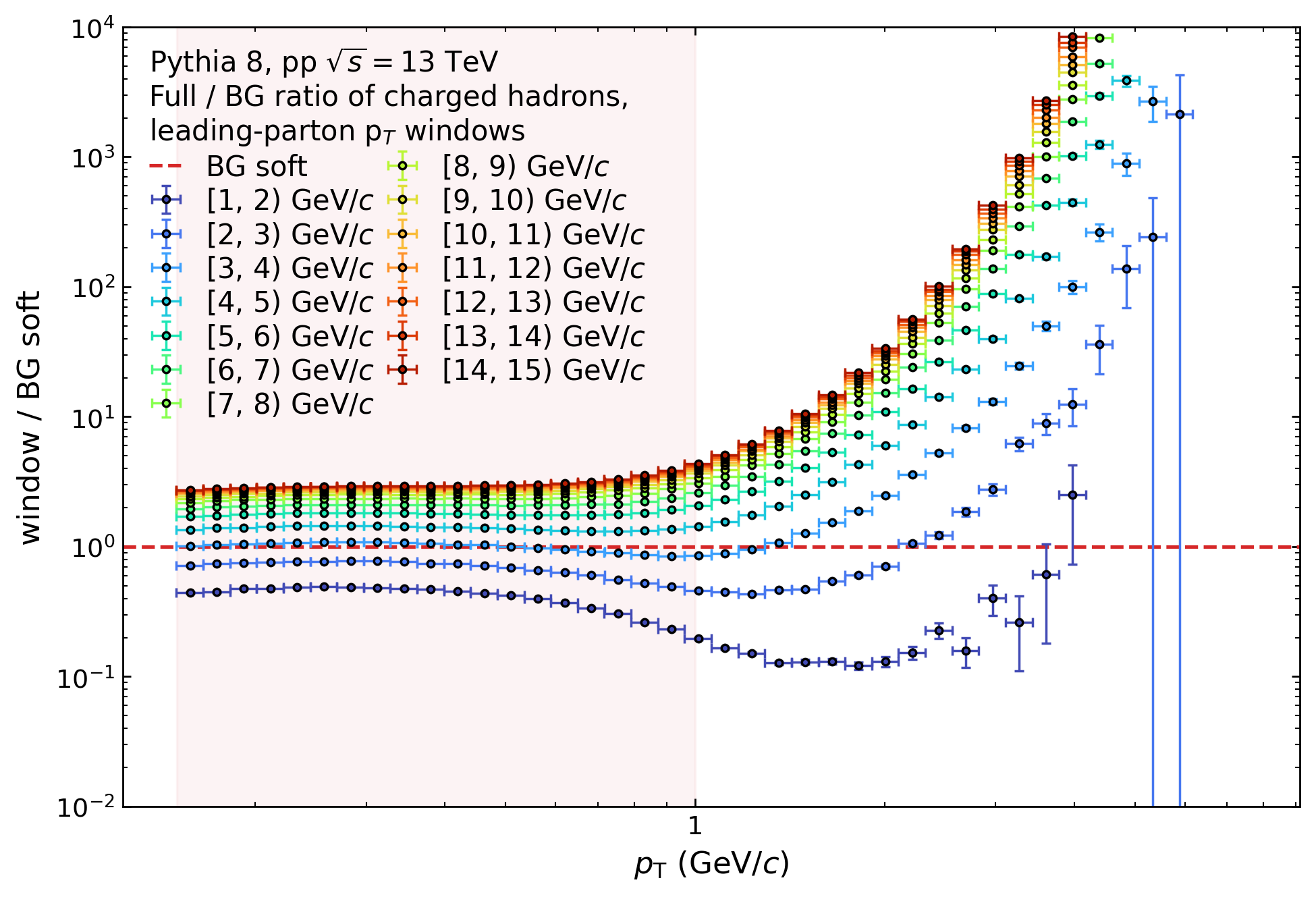}
\caption{Ratio of the full $p_T$ spectrum to the fitted Boltzmann background component in {\sc Pythia}~8 (Monash tune) simulations at $\sqrt{s} = 13$~TeV, shown for all leading-parton $p_T$ event windows. The crossover point where the hard contribution exceeds the soft is stable at $p_T \approx 1.7$~GeV/$c$ across all event-activity windows.}
\label{fig:ratio_to_bg}
\end{figure}

\begin{figure}[h]
\centering
\includegraphics[width=0.48\linewidth]{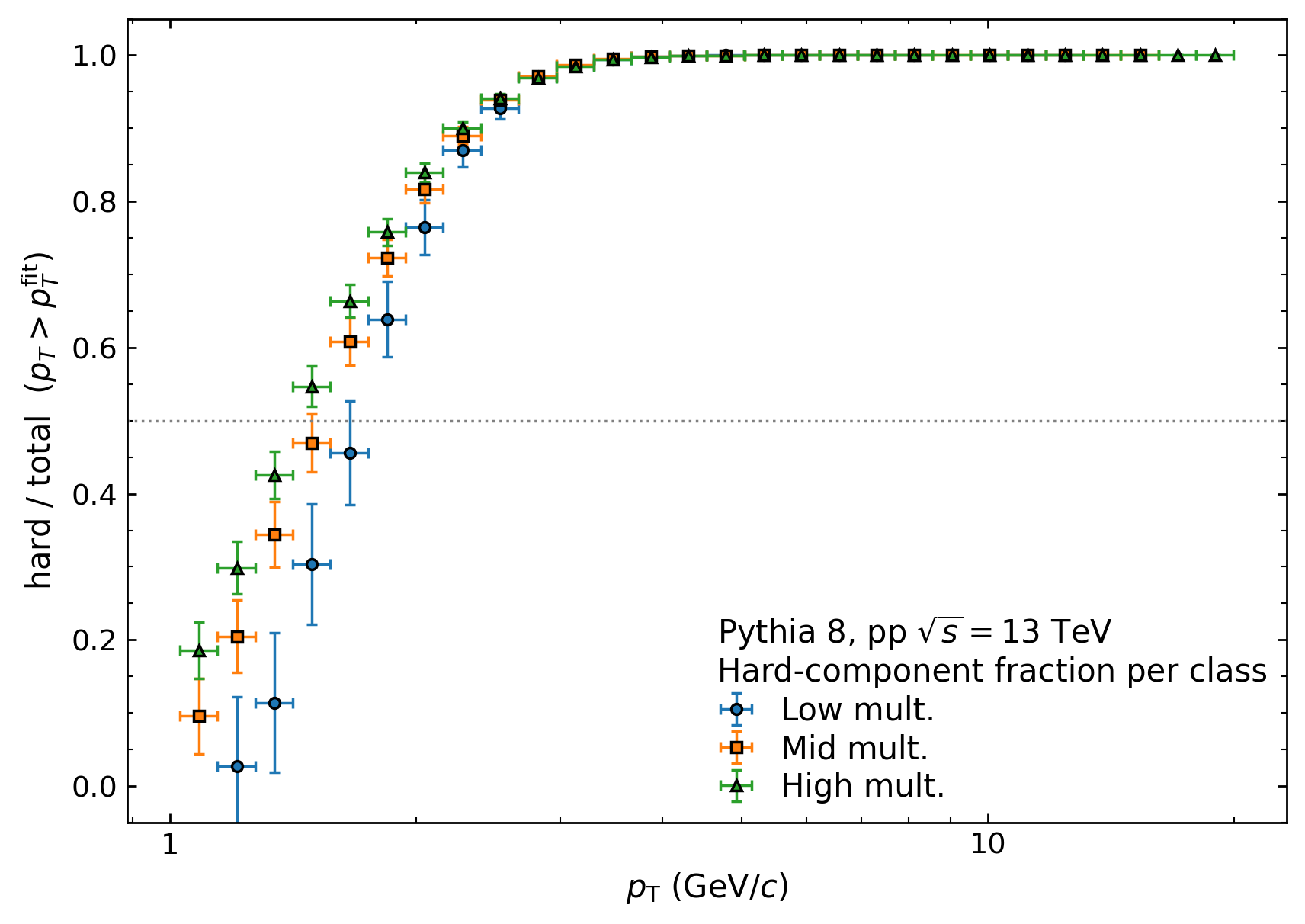}
\caption{Hard-component fraction (ratio of hard-component yield to total yield) as a function of hadron $p_T$, shown for three charged-particle multiplicity classes in {\sc Pythia}~8 (Monash tune) at $\sqrt{s} = 13$~TeV. The fraction rises steeply above $p_T \approx 1.7$~GeV/$c$ in all classes; higher-multiplicity classes show a systematically larger hard fraction, reflecting the growing contribution from harder scatterings at higher event activity.}
\label{fig:hard_fraction}
\end{figure}

To further characterize the decomposition, Fig.~\ref{fig:ratio_to_bg} shows the ratio of the full spectrum to the fitted Boltzmann background for different leading-parton $p_T$ event windows in {\sc Pythia}~8 (version 8.309) simulations at 13~TeV~\cite{Sjostrand:2014zea}, using the Monash tune~\cite{Skands:2014pea}. The Monash tune is the default general-purpose tune of {\sc Pythia}~8, optimized using LHC minimum-bias and underlying-event data at multiple energies. It is the standard reference for minimum-bias $pp$ comparisons at the LHC and was chosen here as the most widely validated tune for this analysis. On Fig.~\ref{fig:ratio_to_bg}, the {\sc Pythia}~8 events are classified by the transverse momentum of the leading parton in the hard $2\to 2$ scattering ($p_T^{\rm hard}$), which serves as a proxy for event activity: windows of increasing $p_T^{\rm hard}$ correspond to events with increasing hard-scattering intensity and, consequently, higher charged-particle multiplicity. This event classification is complementary to the experimental multiplicity classes and allows a controlled study of the multiplicity dependence of the soft and hard components. The crossover from below to above unity marks the $p_T$ scale where the hard fragmentation contribution begins to dominate. This crossover occurs at $p_T \approx 1.7$~GeV/$c$ and is remarkably stable across all event-activity windows. Figure~\ref{fig:hard_fraction} further quantifies this picture by showing the hard-component fraction (ratio of hard-component yield to total yield) as a function of $p_T$, separately for three multiplicity classes (the \textit{Mid multiplicity} denotes the mean charged hadron multiplicity of minimum bias events). The hard fraction rises steeply above $p_T \approx 1.7$~GeV/$c$ and is systematically higher for more active event classes, confirming that harder scatterings produce progressively more particles in this regime, while the shape of the onset is universal.

\begin{figure}[h]
\centering
\includegraphics[width=0.49\linewidth]{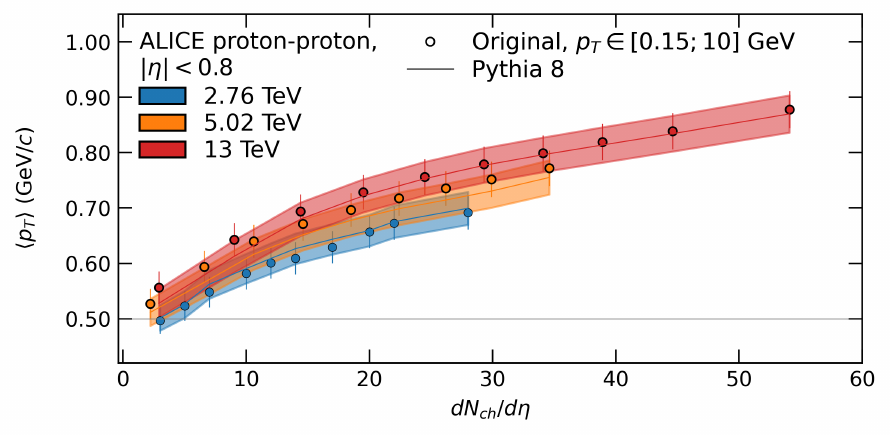}
\includegraphics[width=0.45\linewidth]{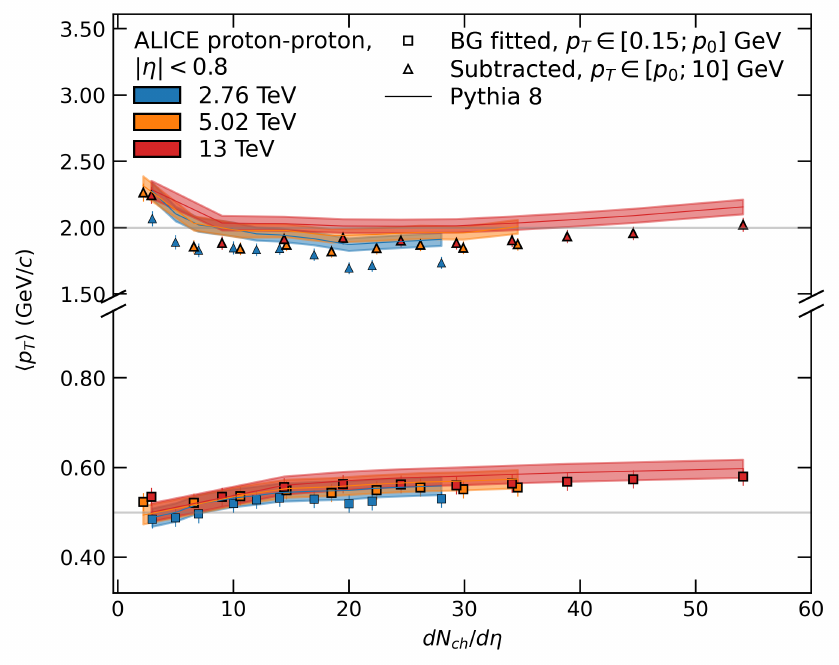}
\caption{
The mean transverse momenta, $\langle p_{\rm T} \rangle$ obtained in the upper panel for the total spectra and in the lower panel the fragmentation and soft contributions from top to bottom, respectively, and as a function of charged particle multiplicity. Data points are taken from ALICE measurements at $\sqrt{s} = 2.76$~TeV, 5.02~TeV, and 13 TeV~\cite{ALICE:2019dfi, ALICE:2015qqj, ALICE:2020nkc}. The lines with uncertainty bands denote {\sc Pythia}~8 predictions. Uncertainty bands on the {\sc Pythia}~8 curves reflect the propagated fit parameter uncertainties.}
\label{fig:meanpT}
\end{figure}

Finally, we extracted the mean transverse momenta, $\langle p_{\rm T} \rangle$ for the soft, hard (fragmentation), and total contributions, as shown in Fig.~\ref{fig:meanpT}. The uncertainties on $\langle p_T \rangle$ for the soft and hard components are propagated from the Boltzmann fit parameter uncertainties ($T \pm \sigma_T$, $A \pm \sigma_A$) and the statistical errors of the data through the subtraction procedure. The plot presents the $\langle p_{\rm T} \rangle$ values as a function of charged particle multiplicity at midrapidity, taken from ALICE measurements at three different LHC energies $\sqrt{s} = 2.76$~TeV, 5.02~TeV, and 13 TeV~\cite{ALICE:2019dfi, ALICE:2015qqj, ALICE:2020nkc}. These results are shown separately for the original full spectra, the Boltzmann-fitted soft part, and the subtracted (hard) component.

One observes that in the upper panel the mean $\langle p_{\rm T} \rangle$ of the total spectra increases monotonically with multiplicity, which is the well-known experimental trend. However, after the soft-hard separation is applied in the lower panel, both individual components exhibit a much flatter dependence. This stability of the soft component across multiplicity further supports our assumption that the exponential part captures a thermal-like contribution with a nearly constant effective temperature. This observation is in line with previous theoretical suggestions linking the exponential slope parameter to $\langle p_{\rm T} \rangle$ and temperature~\cite{Gardim:2024zvi}.

Moreover, the hard component shows only a moderate increase of $\langle p_{\rm T} \rangle$ with multiplicity, reinforcing the idea that fragmentation physics, modelled as a power-law-like tail, becomes more dominant but does not shift dramatically in scale. The flattening observed after subtraction is consistent with the interpretation that the increase of the total $\langle p_{\rm T} \rangle$ with multiplicity is driven by a growing contribution from hard processes rather than by a change in the soft sector.

Fig.~\ref{fig:meanpT} compares the experimental results with {\sc Pythia}~8. The Monte Carlo predictions successfully describe the qualitative features and follow the same trends as the data, both for the total and separated components. We emphasize that this comparison is a consistency check of the fitting and subtraction procedure, not a claim that {\sc Pythia}~8 uniquely validates the decomposition. {\sc Pythia}~8 does not invoke hydrodynamics; instead it generates multiplicity-dependent $p_T$ hardening through MPI, ISR/FSR, and color reconnection (CR). The agreement therefore demonstrates that the observed trends can be reproduced by a microscopic, non-hydrodynamic model.

\begin{figure}[h]
\centering
\includegraphics[width=0.48\linewidth]{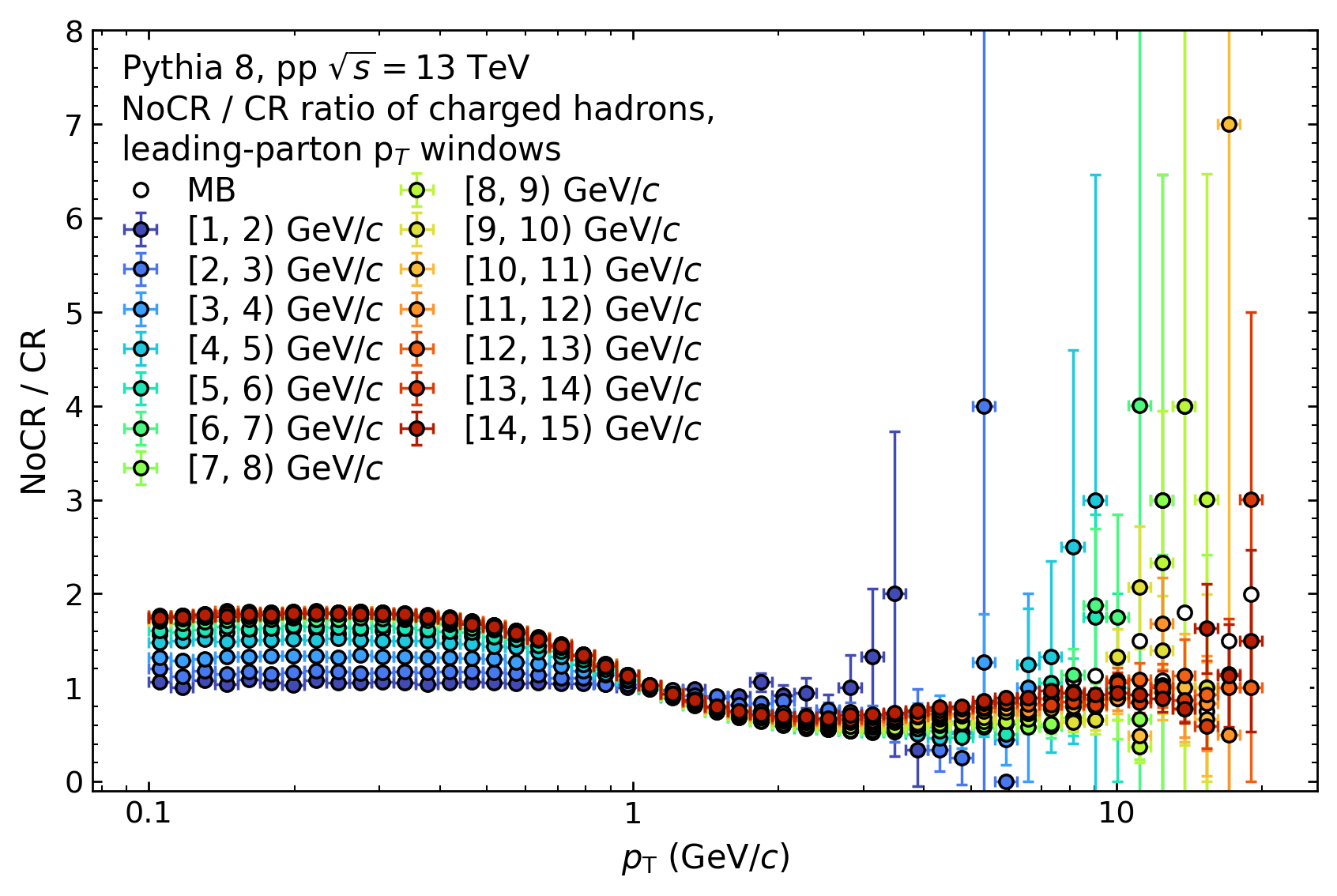}
\caption{Ratio of CR to NoCR $p_T$ spectra in {\sc Pythia}~8 (Monash tune) simulations at $\sqrt{s} = 13$~TeV, shown for all leading-parton $p_T$ event windows. Differences remain at the few-percent level across the full $p_T$ range and all windows. The Boltzmann slope parameter $T$ changes by less than 3\% ($T_{\rm CR} = 0.232 \pm 0.001$~GeV to $T_{\rm noCR} = 0.239 \pm 0.005$~GeV), confirming that the two-component decomposition is robust against color reconnection.}
\label{fig:cr_comparison}
\end{figure}

To assess the role of color reconnection, we have repeated the {\sc Pythia}~8 analysis with CR switched off. The Boltzmann slope parameter $T$ changes by less than 3\% (from $T_{\rm CR} = 0.232 \pm 0.001$~GeV to $T_{\rm noCR} = 0.239 \pm 0.005$~GeV). Figure~\ref{fig:cr_comparison} shows the ratio of CR to NoCR spectra for all leading-parton $p_T$ event windows: differences remain at the few-percent level across the full $p_T$ range and in all windows, confirming that the two-component behaviour is robust under this significant change in the {\sc Pythia}~8 physics settings.

\section{Discussion and summary}

In this work we presented a systematic study of the separation of soft and hard contributions to the transverse momentum spectra of charged particles produced in $pp$ collisions at the LHC. Using ALICE data at $\sqrt{s} = 13$~TeV, the low-$p_{\rm T}$ region of the spectra was fitted with a simple Boltzmann function to describe the soft component, while the remaining high-$p_{\rm T}$ part was identified as the hard (fragmentation) contribution. This procedure was performed in different multiplicity classes and extended to include ALICE measurements at $\sqrt{s} = 2.76$~TeV and 5.02~TeV. 

The analysis procedure is summarized as follows: (i) the Boltzmann function $f(p_T) = A\exp(-p_T/T)$ is fitted to the spectrum in the range $0.15 < p_T < p_0$~GeV/$c$; (ii) $p_0$ is determined per multiplicity class as the value minimizing $|\chi^2/\text{ndf} - 1|$, yielding $p_0 \approx 0.9$--$1.1$~GeV/$c$; (iii) the hard component is obtained by subtracting the Boltzmann curve from the total spectrum for $p_T > p_0$; (iv) $\langle p_T \rangle$ for each component is computed by integrating $p_T \cdot f(p_T)$ over the respective ranges, with uncertainties propagated from the fit parameters.

\begin{figure}[h]
\centering
\includegraphics[width=0.98\linewidth]{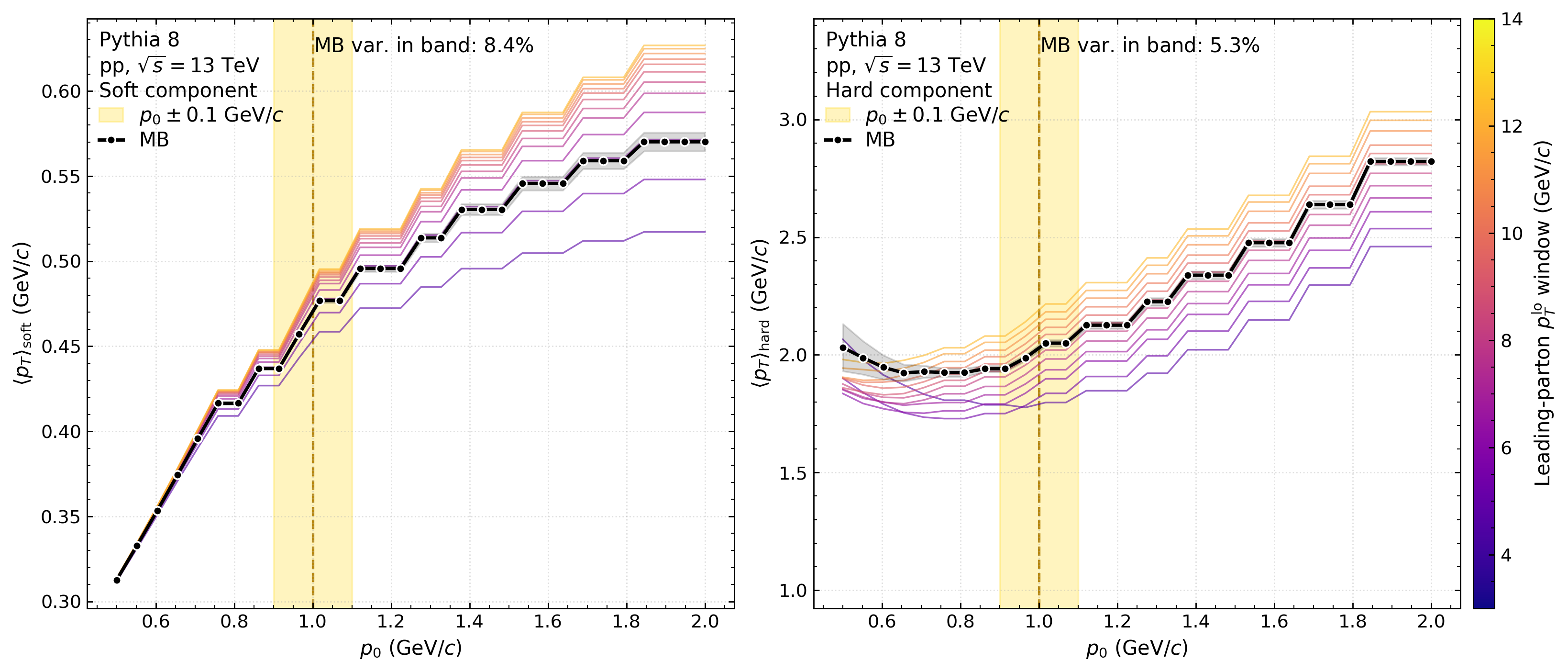}
\caption{$\langle p_T \rangle$ of the soft (left) and hard (right) components as a function of the fit cutoff $p_0$, in {\sc Pythia}~8 (Monash tune) at $\sqrt{s} = 13$~TeV. The minimum-bias result is shown as a black curve; coloured lines show all leading-parton $p_T$ event windows from 3--4~GeV/$c$ upward. The gold band marks $p_0 \pm 0.1$~GeV/$c$ around the nominal value. Within this band, the MB $\langle p_T \rangle_{\rm soft}$ varies by 8.4\% and $\langle p_T \rangle_{\rm hard}$ by 5.3\%, confirming stability of the decomposition.}
\label{fig:p0sensitivity}
\end{figure}

The sensitivity of $\langle p_T \rangle$ to the choice of $p_0$ was assessed by scanning $p_0$ over the range 0.5--2.0~GeV/$c$ for both the minimum-bias spectrum and all leading-parton $p_T$ event windows, as shown in Fig.~\ref{fig:p0sensitivity}. Within $p_0 \pm 0.1$~GeV/$c$ around the nominal value, $\langle p_T \rangle_{\rm soft}$ varies by 8.4\% and $\langle p_T \rangle_{\rm hard}$ by 5.3\% (MB, {\sc Pythia}~8, 13~TeV), confirming that the conclusions are stable with respect to the specific choice of $p_0$.

We found that after the separation both the soft and hard components show a weak dependence on multiplicity in their mean transverse momenta, in contrast to the total $\langle p_{\rm T} \rangle$ which rises with multiplicity. This behaviour is consistent with the interpretation that the increase of the total $\langle p_{\rm T} \rangle$ with multiplicity is driven by a growing contribution from hard processes rather than by a change in the soft sector. The resulting fragmentation tails also retain a stable shape across multiplicity, consistent with the absence of a dominant additional intermediate source such as collective flow within this two-component picture, though we do not claim to uniquely exclude such contributions.

Comparisons with {\sc Pythia}~8 simulations using the Monash tune show consistent trends for both the total and separated components. The comparison with {\sc Pythia}~8 without color reconnection further confirms that the two-component behaviour is not an artifact of CR effects. A systematic study of MPI contributions is left for future work.

The results obtained with a two-component contribution to the spectra, in agreement with QCD expectations, represent an important input to the ongoing discussions about whether $pp$ collisions can indeed be described by a two-component process. It is important to recall that hydrodynamical interpretations based on BW or similar models may involve fit regions where hard-fragmentation contributions are non-negligible, making the interpretation of extracted collective parameters complicated~\cite{Florkowski:2016xbk, Mazeliauskas:2019ifr}. We note, however, that demonstrating consistency with the two-component picture does not by itself rule out hydrodynamical descriptions; both frameworks may describe the same data with comparable quality, since they parametrize overlapping physical effects in different ways.

Our results provide evidence consistent with a two-component description, consisting of a soft Boltzmann-like and a hard QCD-like part, advocated by theorists before the advent of heavy-ion experiments, and more recently by T.~Trainor, as a viable and physically motivated alternative description of $pp$ spectra.

\section*{Acknowledgments}
The authors wish to acknowledge insightful discussions with Prof. T.~Trainor. The research was supported by the Hungarian National Research, Development and Innovation Office (NKFIH) under the contract numbers NKKP ADVANCED\_25-153456, 2025-1.1.5-NEMZ\_KI-2025-00005, 2025-1.1.5-NEMZ\_KI-2025-00013, NKFIH NEMZ\_KI-2022-00058 and 2024-1.2.5-TET-2024-00022, the FuSe COST Action CA-24101 as well as 
the Wigner Scientific Computing Laboratory and the HUN-REN Wigner Cloud. Support for this work has been received from the Mexican Secretary of Science, Humanities, Technology and Innovation SECIHTI under Grant No. CBF-2025-I-173, and from DGAPA--UNAM under the Grant PAPIIT No. IA103626.

\bibliography{Softhard_ref}

\end{document}